\documentclass[dvipdfmx,11pt]{raa}
\usepackage{amssymb}
\usepackage{amsmath}
\usepackage{graphicx,times,booktabs,multirow}
\usepackage{subfig}
\usepackage{graphics}

\begin{document}

\title{On the DB gap of white dwarf evolution: effects of hydrogen mass fraction and
convective overshooting
}

   \author{Jie Su
      \inst{1, 2}
   \and Yan Li
      \inst{1}
   }

   \institute{NAOC / Yunnan Observatory, Chinese Academy of Sciences,
             Kunming 650011, China. {\it sujie@ynao.ac.cn, ly@ynao.ac.cn}\\
    \and Graduate School of the Chinese Academy of Sciences, Beijing 100049,
    China.\\
\vs \no
   {\small Received 2009 Sep. 18; accepted 2009 Nov. 12}    
    }

\abstract{ We investigate the spectral evolution of white dwarfs by
considering the effects of hydrogen mass in the atmosphere and
convective overshooting above the convection zone. Our numerical
results show that white dwarfs with $M_{\rm
H}\sim10^{-16}~M_{\odot}$ show DA spectral type between
$46,000\lesssim T_{\rm eff}\lesssim26,000~{\rm K}$ and DO or DB
spectral type may appears on either side of this temperature range.
White dwarfs with $M_{\rm H}\sim10^{-15}~M_{\odot}$ appear as DA
stars until they cool to $T_{\rm eff}\sim31,000~{\rm K}$, from then
on they will evolve into DB white dwarfs as a result of convective
mixing. If $M_{\rm H}$ in the white dwarfs more than
$10^{-14}~M_{\odot}$, the convective mixing will not occur when
$T_{\rm eff}>20,000~{\rm K}$, thus these white dwarfs always appear
as DA stars. White dwarfs within the temperature range
$46,000\lesssim T_{\rm eff}\lesssim31,000~{\rm K}$ always show DA
spectral type, which coincides with the DB gap. We notice the
importance of the convective overshooting and suggest that the
overshooting length should be proportional to the thickness of the
convection zone to better fit the observations.
 }
   \authorrunning{J. Su \& Y. Li}
   \titlerunning{On the DB gap of white dwarf evolution}

   \maketitle

\section{Introduction}
\label{sect:intro}

White dwarfs can be classified into several spectral types in terms
of the spectral characteristics. The current classification system
was introduced by Sion et al. (\cite{Sc}) and has been modified
several times. In this system, the spectral type of a white dwarf is
denoted by a letter D plus another letter indicating its spectral
characteristics. Sometimes, a suffix is added to indicate some other
features (polarization, magnetic field, pulse, etc.). Table
\ref{spec} lists a spectral classification scheme from McCook \&
Sion (\cite{Ma}).

\begin{table}[ht]
\begin{center}
\caption{White dwarf spectral types}
\begin{tabular}{p{1pt} l l}
      \toprule
      \multicolumn{2}{l}{Spectral Type}
        & \multicolumn{1}{c}{Characteristics}
         \\
      \midrule
       DA && Only Balmer lines; no He I or metals present \\
       DB && He I lines; no H or metals present \\
       DC && Continuous spectrum, no lines deeper than $5\%$ in any part of the electromagnetic spectrum \\
       DO && He II strong; He I or H present \\
       DZ && Metal lines only; no H or He lines \\
       DQ && Carbon features, either atomic or molecular in any part of the electromagnetic spectrum \\
       P & (suffix) & Magnetic white dwarfs with detectable polarization \\
       H & (suffix) & Magnetic white dwarfs without detectable polarization \\
       X & (suffix) & Peculiar or unclassifiable spectrum \\
       E & (suffix) & Emission lines are present \\
       ? & (suffix) & Uncertain assigned classification; a colon (:) may also be used \\
       V & (suffix) & Optional symbol to denote variability \\
      \bottomrule
\end{tabular}
\label{spec}
\end{center}
\end{table}

Most of white dwarfs are of DA type which have hydrogen-dominated atmospheres.
They are found at all effective temperatures from $170,000~{\rm K}$ down to
about $4,500~{\rm K}$ (Kurtz et al. \cite{Kb}). DA white dwarfs occupy the vast
majority (about $75\%$) of all known white dwarfs.

About $25\%$ of the observed white dwarfs have helium-dominated atmospheres which
can be divided further into two spectral types. The DO white dwarfs are found
between approximately $100,000$ to $45,000~{\rm K}$, their outer atmospheres being
dominated by singly ionized helium (He II). The DB white dwarfs are found between
approximately $30,000$ to $12,000~{\rm K}$, their outer atmospheres being
dominated by neutral helium (He I).

There is an interesting fact that few white dwarfs with
helium-dominated atmospheres (DO or DB type) are found in the
effective temperature range of $45,000\lesssim T_{\rm eff} \lesssim
30,000~{\rm K}$. This is the so-called DB gap. The reason for this
phenomenon is not clear. On one side, there is strong gravitational
field in the white dwarf, which may cause a stratified atmosphere.
The so-called gravitational settling effect lets the light element
(hydrogen) floating to the stellar surface and the heavy element
(helium) sinking to the bottom of the stellar envelope. The typical
mass fractions of hydrogen and helium in the white dwarfs are
$M_{\rm H}/M_{\rm tot}\lesssim 10^{-4}$ and $M_{\rm He}/M_{\rm
tot}\lesssim 10^{-2}$, which are the mass threshold for residual
nuclear burning (Tremblay \& Bergeron \cite{T}). If there is no
mixing between hydrogen and helium, we will expect all white dwarfs
to be DA stars. On the other side, however, many observational data
show that $M_{\rm H}$ may be significantly lower than the typical
value. The existence of a large number of non-DA white dwarfs
indicates that some physical mechanisms are competing with the
gravitational settling and make the spectral type of some DA white
dwarfs changed. The DB gap is suspected to be due to the competition
between the gravitational settling and convective mixing.

Fontaine \& Wesemael (\cite{Fb}) proposed that when a white dwarf starts cooling
from the hot PG $1159$ type star, hydrogen is mixed within the outer helium envelope
and the white dwarf shows the DO spectral type. After that, hydrogen gradually
floats to the stellar surface in the strong gravitational field. When the DO star
cools to $T_{\rm eff}\sim45,000~{\rm K}$, hydrogen has accumulated enough at the
surface, and the white dwarf is turned into a DA star. They supposed further that
as soon as the DA star cools down to $T_{\rm eff}\sim30,000~{\rm K}$, the He I/II
ionization zone in the stellar envelope becomes convective. The convective motion
may penetrate into the top hydrogen layer and mix the hydrogen atmosphere into the
helium envelope, the less abundant hydrogen being overwhelmed by the more abundant
helium and becoming undetectable. So the DA star then appears as a DB white dwarf.

Shibahashi (\cite{Sa}, \cite{Sb}) proposed a different scenario for
the DB gap. During the early stage of a white dwarf's evolution, the
convection in the He II/III ionization zone mixes the hydrogen layer
and results in a helium-dominated atmosphere, and the white dwarf
appears as a DO star. When the star cools down to around $T_{\rm
eff}\sim45,000~{\rm K}$, the He II/III ionization zone becomes deep
enough that convection disappears in the DO star's atmosphere, so
hydrogen floating to the surface and then the white dwarf being
transformed into a DA star. When the white dwarf cools further down
to $T_{\rm eff}\sim30,000~{\rm K}$, the He I/II ionization zone
generates a convection zone again, as Fontaine \& Wesemael's
proposal, a similar mixing between H and He occurs and the white
dwarf appears as a DB star.

The difference between the two scenarios of the DB gap is the time
scale of the gravitational settling. In Fontaine \& Wesemael's
assumption, the settling process happens slowly as the cooling of
the white dwarf. But in Shibahashi's assumption, it happens quickly
as soon as the convection is turned off. However, the recent data of
the Sloan Digital Sky Survey (SDSS) suggest that several DB white
dwarfs do appear in the DB gap (Eisenstein et al. \cite{E}). These
facts imply that the formation mechanism of the DB gap are not clear
and more works should be done.

In the present paper, we calculate a series of white dwarf
evolutionary models to investigate the spectral evolution of white
dwarfs caused by the convective mixing. The details of model
calculations and input physics are presented in Section 2. In
Section 3 we discuss the results of our numerical models.
Conclusions are summarized in Section 4.

\section{model details and input physics}
\label{sect:mod}

We have used a modified version of the White Dwarf Evolution Code
(WDEC), which was originally described by Martin Schwarzschild to
simulate the evolution of the white dwarf. Some details of the WDEC
has been described in Lamb \& Van Horn (\cite{Lb}) and Wood
(\cite{W}). Here, we only present some summaries of the input
physics in our models.

The equation of state (EOS) used in the present calculations is composed of two
parts which apply to different regions. The first part of the EOS is used for the
degenerate, completely ionized interior of the white dwarf. In this region, we use
the EOS tables provided by Lamb (\cite{La}). For a given chemical composition the
needed values are obtained by two-dimension, four-point Aitken-Lagrange interpolation
in terms of variables $\lg P$ and $\lg T$. For a specific C/O mixture, they are
obtained by interpolation between the carbon and oxygen tables using the additive
volume technique of Fontaine et al. (\cite{Fa}). The second part of the EOS is used
for the partial ionized envelope where non-ideal effect is important. We use the
Saumon et al. (\cite{Saumon95}) EOS for hydrogen and helium mixtures. The new EOS
include some new physical treatments of partial ionization caused by pressure and
temperature. Mixtures of hydrogen and helium are also obtained by the additive
volume technique.

The total opacity ($\kappa$) is given by
\begin{equation}
\dfrac{1}{\kappa}=\dfrac{1}{\kappa_{\rm r}}+\dfrac{1}{\kappa_{\rm c}}~,
\end{equation}
where $\kappa_{\rm r}$ is the radiative opacity and $\kappa_{\rm c}$ is the
conductive opacity. We use the OPAL radiative opacities in our calculations.
The new tables include some new physical factors, e.g. the L-S coupling effect of
iron atoms. The conductive opacities consist of two parts which are from Itoh et
al. (\cite{Ia},~\cite{Mb}) and Hubbard \& Lampe (\cite{H}). In the actual
calculations, we use Itoh et al. opacities only in $\lg\rho\ge1.8$ region and
Hubbard \& Lampe opacities in $\lg\rho\le1.5$ region, in the range
of $1.5<\lg\rho<1.8$ linear interpolation being performed.

The rates of neutrino energy loss used in our calculations are provided by Itoh
and his collaborators. The rates of neutrino energy loss due to pair, photo, plasma
and bremsstrahlung processes are from Itoh et al. (\cite{Ib}),
and the rate of recombination neutrino is from Kohyama et al. (\cite{Ka}).

\begin{figure}[ht]
 \begin{center}
  \includegraphics[scale=1.5]{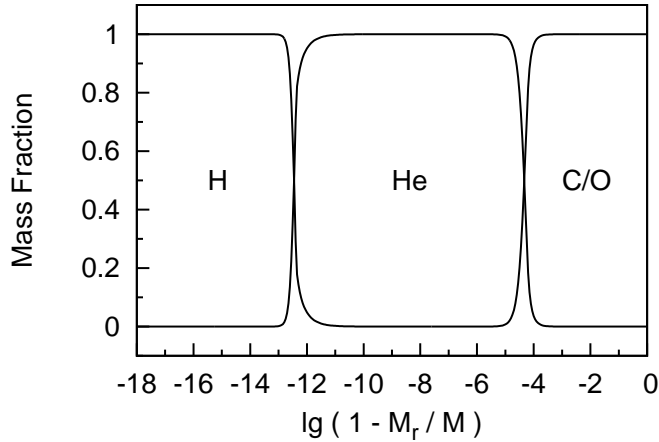}
   \caption{Approximations to the diffusive equilibrium profiles.}
   \label{dpf}
 \end{center}
\end{figure}

The high surface gravity of the white dwarf leads to gravitational segregation of
the elements in the stellar envelope, and thus models of white dwarf must include
compositionally stratified envelopes. In our calculations, we adopt approximations
of the equilibrium diffusion profiles introduced by Wood (\cite{W}) (see Figure \ref{dpf}).
Our calculations do not include the impact of convective mixing on the H/He profile.
Although this may not be correct in details, we still expect it to be a reasonable
approximation, because the abundance of hydrogen in the mixing region is several
orders of magnitude less than the helium abundance.

We use the standard mixing-length theory of B\"ohm-Vitense (\cite{B}) to deal with
the convection. We set the mixing-length $l$ to be equal to one local pressure scale
height, i.e.,
\begin{equation}
l=H_{\rm P}=-\cfrac{dr}{d\ln P}=\cfrac{P}{\rho g}~.
\end{equation}
Boundaries of the convection zones are determined by the Schwarzschild criterion. And 
we set the integration step to be equal to $H_{\rm P}/8$ in our calculations.

\section{evolutionary results}
\label{sect:res}

We have computed a series of white dwarf evolutionary models with
mass $M=0.6~M_{\odot}$, which is the typical mass of white dwarf. In
order for a DA white dwarf to change its surface chemical
composition as a result of the convective mixing, it ought to have a
very thin hydrogen layer, so the hydrogen mass of the model is
supposed to vary between $10^{-16}$ to $10^{-14}~M_{\odot}$. The
helium mass in the computed model envelope is fixed to be
$5.0\times10^{-5}~M_{\odot}$. We assume that the heavy elements 
have sunk during the early phase of the cooling process to the white 
dwarf's interior due to the so-called gravitational settling effect. 
As a result the envelope of the white dwarf is only composed of hydrogen 
and helium, and the metallicity in the envelope is assumed to be $Z=0$. 
All models are evolved from $T_{\rm eff}\sim 90,000~{\rm K}$ down to 
$T_{\rm eff}=10,000~{\rm K}$. In this section we present the results 
of our calculations.

\subsection{Importance of the convective overshooting}
\label{subsect:ovs}

\begin{figure}[ht]
\centering
\subfloat[]{
    \includegraphics[scale=0.75]{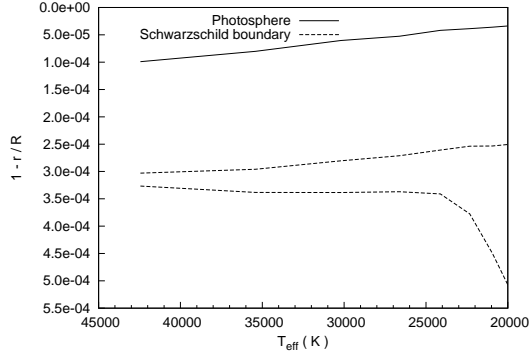}
}
\subfloat[]{
    \includegraphics[scale=0.75]{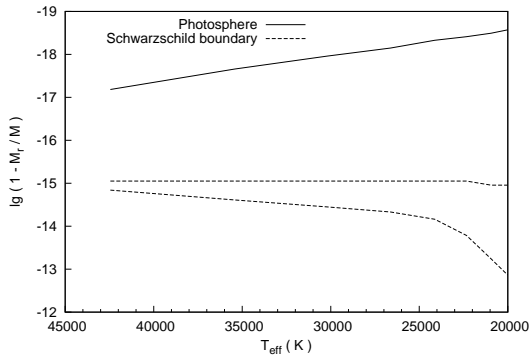}
}
\caption{\label{crm}Location of the convection zone.}
\end{figure}

We have examined the development of convection zone in our white dwarf models.
Figure \ref{crm} shows the convection zone of a DA model
with $M_{\rm H}=10^{-15}~M_{\odot}$ evolving with the decreasing effective temperature.
The horizontal axis is the effective temperature which denotes the evolutionary
sequence, and the vertical axis denotes the location within the envelope of the model.
We use $1-r/R$ as the vertical axis scale in Figure \ref{crm} (a) and
use $\lg(1-M_r/M)$ in Figure \ref{crm} (b). The solid line corresponds to the location
of the model's photosphere ($\tau =1$) and the dashed lines correspond to the
Schwarzschild boundaries of the convection zone.

\begin{figure}[ht]
\begin{center}
\includegraphics{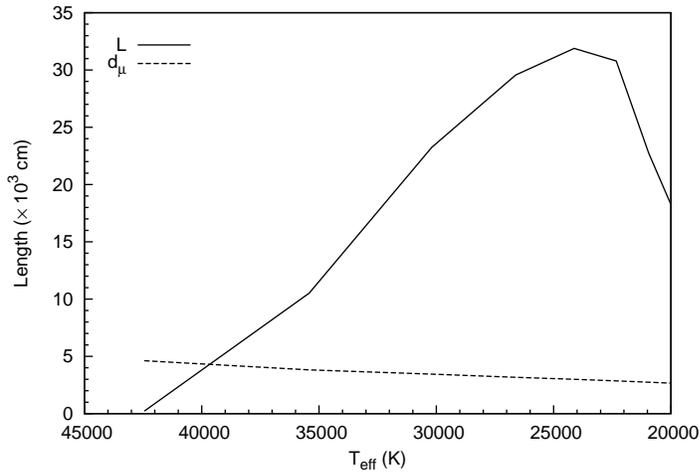}
\caption{$L$ and $d_{\rm \mu}$ versus $T_{\rm eff}$.} 
\label{lvd}
\end{center}
\end{figure}

\begin{figure}[ht]
\begin{center}
\includegraphics{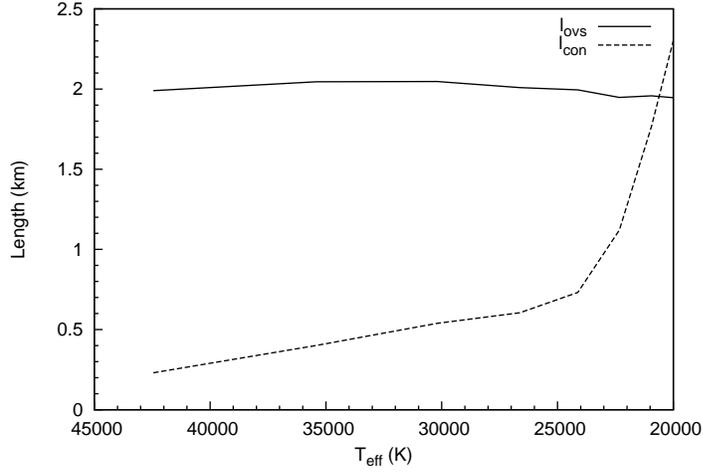}
\caption{The evolution of the length of overshooting region and that of the
convection zone. $l_{\rm ovs}$ is the length of the overshooting region
and $l_{\rm con}$ the length of the convection zone.}
\label{lvl}
\end{center}
\end{figure}

\begin{figure}[ht]
\begin{center}
\includegraphics{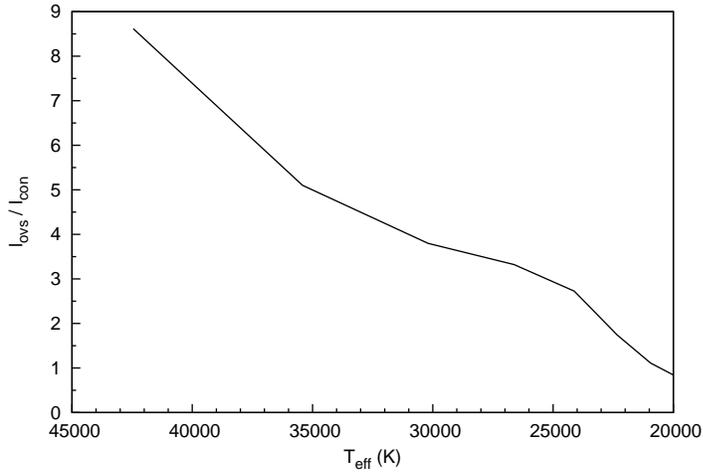}
\caption{The ratio of $l_{\rm ovs}$ to $l_{\rm con}$ varies with $T_{\rm eff}$.}
\label{rol}
\end{center}
\end{figure}

It can be found that convection occurs completely within the helium
layer which is located under the photosphere which can be regarded
as the innermost point visible to us in the white dwarf. Just above 
the He convection zone there is a thin layer with a mean molecular 
weight gradient (the so-called $\mu$-barrier). In the case of very thin 
hydrogen envelope ($M_{\rm H}<10^{-15}~M_{\odot}$), Vauclair \& Reisse 
(\cite{VR}) have argued that the $\mu$-barrier provides extra 
buoyancy to restrict the upper boundary of the He convection zone 
just below it. However, at the upper boundary which is determined 
by the convective stability criterion (at the $\mu$-barrier, the Ledoux criterion 
$\nabla_{\rm rad}\ge \nabla_{\rm ad}+d\ln\mu/d\ln P$ is adopted), 
the convective motion does not stop but moves upward further due to inertia. 
In this sense, the fluid parcels with kinetic energy may penetrate into the 
$\mu$-barrier until their velocity drop to zero. We perform a rough calculation, 
supposing that the velocities of the fluid parcels is equal to the mean flow 
velocity ($\bar{v}$) in the convection zone. When the fluid parcels enter 
the $\mu$-barrier, a force $F$ (the resultant force of gravity and buoyancy, 
neglecting the viscous force) acts on them and reduces the velocity. 
The kinetic energy of the fluid parcels $E_{\rm K}\sim\bar{v}^2/2$.  
When the fluid parcels deplete all the kinetic energy, they can 
move beyond the convective boundary a length 
$L\sim E_{\rm K}/F\sim\bar{v}^2/2F$. As shown 
in Fig. \ref{lvd}, the thickness of the $\mu$-barrier $d_{\rm \mu}$ is 
around $3\times 10^3~{\rm cm}$ and $L>d_{\rm \mu}$ when 
$T_{\rm eff}<40,000~{\rm K}$, and the maximum value of $L$ is about 
$3\times 10^4~{\rm cm}$ which is one order of magnitude greater 
than $d_{\rm \mu}$. Therefore, it seems reasonable to believe that 
the convective overshooting can go cross the $\mu$-barrier. Once He 
penetrates into the $\mu$-barrier, an effective mixing will 
flatten the composition gradient and weaken the $\mu$-barrier. As a
result, the upper boundary of the He convection zone will thus be able 
to extend outward. 
We expect that the convective motion can extend upward to the stellar
surface, or at least, to the photosphere in order to mix the upper
hydrogen layer and to ensure that helium can be observed. Thus, the
role of convective overshooting appears to be decisive. We suppose
that the convective overshooting is able to reach the photosphere.
This suggestion let us set the distance from the photosphere down to
the top of the Schwarzschild boundary as the minimum length of the
overshooting region ($l_{\rm ovs}$). The geometrical length between
the two Schwarzschild boundaries is regarded as the length of the
convection zone ($l_{\rm con}$). These two lengths vary with the
evolution of the effective temperature as shown in Figure \ref{lvl}.
We also compare $l_{\rm ovs}$ with $l_{\rm con}$, the ratio given in
Figure \ref{rol}.

\begin{figure}[ht]
\begin{center}
\includegraphics{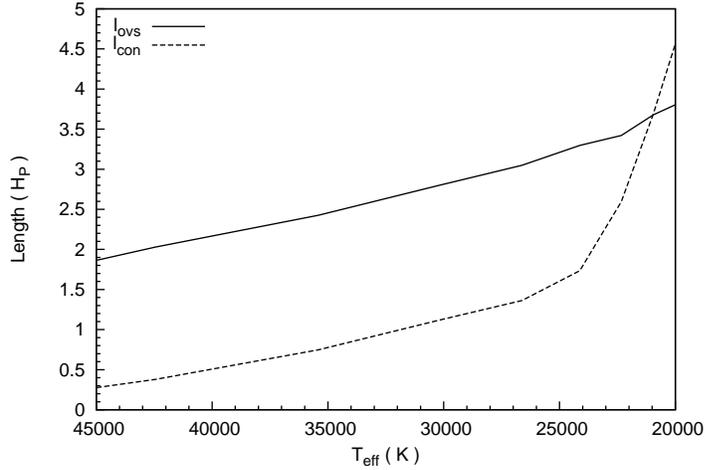}
\caption{Similar to Figure \ref{lvl}, but the two lengths are
expressed in unit of $H_{\rm P}$.}
\label{lhp}
\end{center}
\end{figure}

As shown in the Figure \ref{lvl}, $l_{\rm ovs}$ keeps almost a constant
(about $2~{\rm km}$), while $l_{\rm con}$ increases slowly to a few hundred
meters during a long evolutionary time scale. That is to say, the convective motion
must overshoot to a distance which is several times thicker than the convection zone
itself. Figure \ref{lhp} is similar to Figure \ref{lvl} but $l_{\rm ovs}$
and $l_{\rm con}$ are expressed in unit of the local pressure scale
height ($H_{\rm P}$). It is shown that $l_{\rm ovs}$ is no more than $4H_{\rm P}$.

\begin{figure}[ht]
\begin{center}
\includegraphics{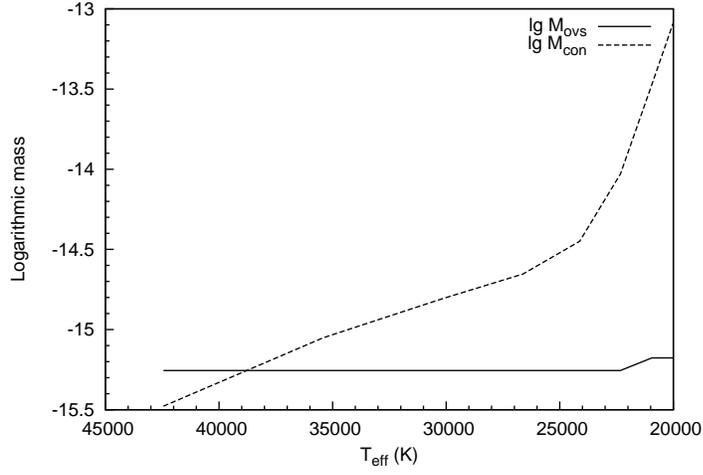}
\caption{Masses of overshooting region and convection zone vary with $T_{\rm eff}$.
$M_{\rm ovs}$ is the mass of the overshooting region and $M_{\rm con}$ is the
mass of the convection zone.}
\label{mvm}
\end{center}
\end{figure}

\begin{figure}[ht]
\begin{center}
\includegraphics{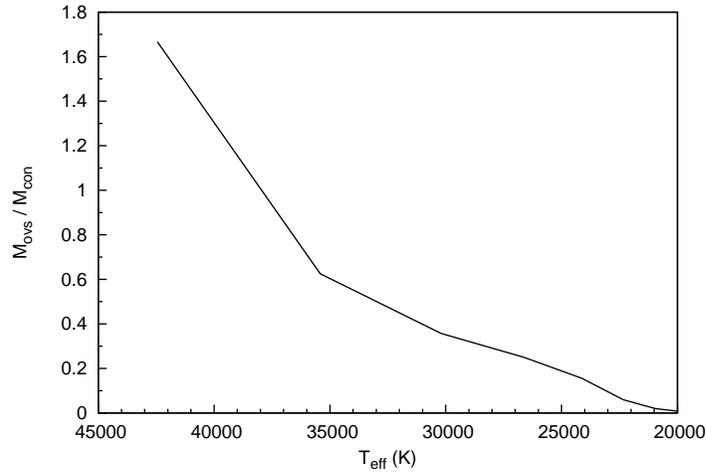}
\caption{The ratio of $M_{\rm ovs}$ to $M_{\rm con}$ varies with $T_{\rm eff}$.}
\label{rom}
\end{center}
\end{figure}

We discuss from another side of view the extension of convective
overshooting that concerns the masses in the convection and
overshooting regions. We denote $M_{\rm ovs}$ the mass within the
overshooting region and $M_{\rm con}$ the mass of the convection
zone. Figure \ref{mvm} shows $\lg M_{\rm ovs}$ and $\lg M_{\rm con}$
vary with the effective temperature, respectively. Figure \ref{rom}
shows the variation of the ratio of $M_{\rm ovs}$ to $M_{\rm con}$.
It can be seen that $M_{\rm ovs}/M_{\rm con}$ decreases rapidly with
$T_{\rm eff}$ and $M_{\rm ovs}$ accounts only for a small fraction
of $M_{\rm con}$ for the models with relatively low $T_{\rm eff}$.
The convection zone is thickening in the evolutionary process and
causes a rapid increase in $M_{\rm con}$. At $T_{\rm
eff}\sim38,000~{\rm K}$, $M_{\rm con}$ is greater than $M_{\rm
ovs}$. It is interesting to note, that although $l_{\rm ovs}$ is
considerably large than $l_{\rm con}$, the matter in the
overshooting region has a low density comparing with the dense,
turbulent convection zone. For example, when the white dwarf model
cools down to $T_{\rm eff}\sim30,000~{\rm K}$, the ratio of $l_{\rm
ovs}$ to $l_{\rm con}$ is less than $3.8$ and $M_{\rm ovs}$ is only
about $1/3$ of $M_{\rm con}$. Therefore, we may reasonably believe
that the convective motion can extend to the photosphere by force of
the convective overshooting.

\subsection{Determination of the transition temperature}
\label{subsect:dtr}

If the convection zone in the helium envelope of a DA white dwarf
can extend upward to the photosphere due to the convective
overshooting, the hydrogen atmosphere will be mixed into the
convective helium layer and the white dwarf will have the
opportunity to change its apparent chemical composition, in other
words, to transform into a DB star. In this section, we will discuss
when such transformation occurs.

We assume that the convective mixing region includes the overshooting region and the
convection zone, in which hydrogen and helium are homogeneously mixed. The mass of
the mixing zone is denoted as $M_{\rm mix}$, and the mass of hydrogen in the mixing
zone is denoted as $M_{\rm Hmix}$. The remainder is helium whose mass is equal
to $M_{\rm mix}-M_{\rm Hmix}$. When the mass of helium exceeds the mass of hydrogen
in the mixing zone, i.e.
\begin{equation}
  M_{\rm mix}-M_{\rm Hmix}>M_{\rm Hmix}~,
\end{equation}
that is,
\begin{equation}
  M_{\rm mix}>2M_{\rm Hmix}~,
\label{eqtr}
\end{equation}
hydrogen will be overwhelmed by helium and we assume that the
transformation of the spectral type will occur. We use $M_{\rm
mix}=2M_{\rm Hmix}$ as a critical condition for a DA star evolving
into a DB one. The effective temperature at this critical point is
called the transition temperature.

\subsection{Discussions of the calculation results}
\label{subsect:dis}

\begin{figure}[ht]
\begin{center}
\includegraphics{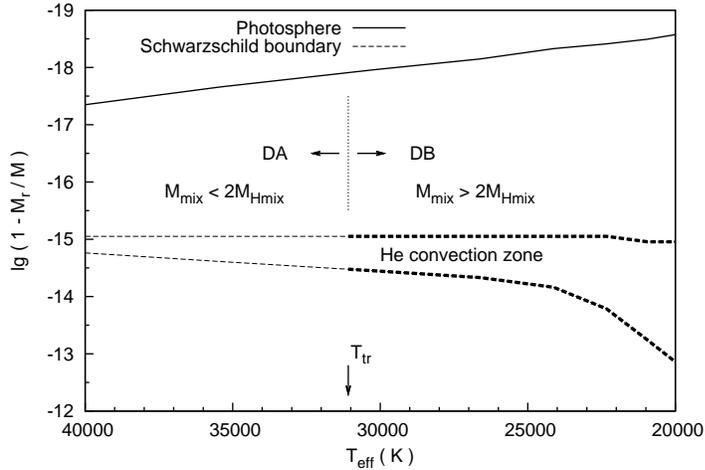}
\caption{A schematic representation of the spectral evolution of a white dwarf model
with $M_{\rm H}=1.0\times 10^{-15}~M_{\odot}$.}
\label{h15}
\end{center}
\end{figure}

We have computed a series model of a DA white dwarf with $M_{\rm
H}=1.0\times 10^{-15}~M_{\odot}$. The convection zone varies with
the effective temperature as shown in Figure \ref{h15}, in which we
use $\lg(1-M_r/M)$ to indicate the location of the convection zone.
The convective motion appears just below the H/He interface
($\lg(1-M_r/M)\approx-15$) and extends into the stellar interior
when the white dwarf model cools down. When the convective mixing
zone becomes thick enough, Eq. (\ref{eqtr}) is satisfied (indicated
by the thick dashed line). It can be found that the transition
temperature ($T_{\rm tr}$) of this model is about $31,000~{\rm K}$.
If the convective overshooting can reach to the photosphere at this
temperature, we will observe that the white dwarf evolves into a DB
star. According to the discussions in Section \ref{subsect:ovs}, the
required length of the convective overshooting is about $3H_{\rm P}$.

Other series of white dwarf evolution models we have computed show
that the thicker the hydrogen layer is, the lower the transition
temperature will be. Table \ref{temptr} lists the value of $T_{\rm
tr}$ of our models with different $M_{\rm H}$. It can be noticed
that the transition temperature of the model with $M_{\rm
H}=1.0\times10^{-14}~M_{\odot}$ is below $20,000~{\rm K}$. It is
expected therefore that models with the hydrogen layers heavier than
$10^{-14}~M_{\odot}$ must have $T_{\rm tr}$ lower than $18,000~{\rm
K}$. So a DA white dwarf with $M_{\rm H}>10^{-15}~M_{\odot}$ may
have the opportunity to change its spectral type when it cools much
below $T_{\rm eff}\sim30,000~{\rm K}$. Moreover, greater $M_{\rm H}$
also requires stronger convective overshooting to bring helium to
the stellar surface.

\begin{table}[ht]
\begin{center}
\caption{Transition temperatures for different white dwarf models}
\begin{tabular}{c|c|c|c|c|c}
      \hline
       \multirow{2}{*}{$M_{\rm H}~(M_{\odot})$} & \multicolumn{2}{c|}{$T_{\rm tr}~(~{\rm K})$} 
      & \multirow{2}{*}{$M_{\rm H}~(M_{\odot})$} & \multicolumn{2}{c}{$T_{\rm tr}~(~{\rm K})$} \\
      \cline{2-3}\cline{5-6}
       & $l=H_{\rm P}$ & $l=2H_{\rm P}$ && $l=H_{\rm P}$ & $l=2H_{\rm P}$ \\
      \hline
       $1.0\times10^{-15}$ & $31,084$ & $31,084$ & $1.8\times10^{-15}$ & $25,191$ & $26,222$ \\
       $1.1\times10^{-15}$ & $29,349$ & $29,769$ & $2.0\times10^{-15}$ & $24,001$ & $26,209$ \\
       $1.2\times10^{-15}$ & $27,920$ & $29,006$ & $3.0\times10^{-15}$ & $22,112$ & $24,004$ \\
       $1.4\times10^{-15}$ & $27,509$ & $28,456$ & $4.0\times10^{-15}$ & $20,785$ & $22,799$ \\
       $1.5\times10^{-15}$ & $27,257$ & $28,240$ & $5.0\times10^{-15}$ & $19,966$ & $22,477$ \\
       $1.6\times10^{-15}$ & $26,360$ & $27,466$ & $1.0\times10^{-14}$ & $18,460$ & $19,802$ \\
      \hline
\end{tabular}
\label{temptr}
\end{center}
\end{table}

\begin{figure}[ht]
\centering
\subfloat[]{
    \includegraphics[scale=0.75]{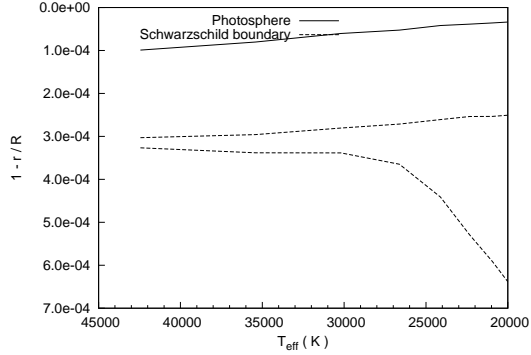}
}
\subfloat[]{
    \includegraphics[scale=0.75]{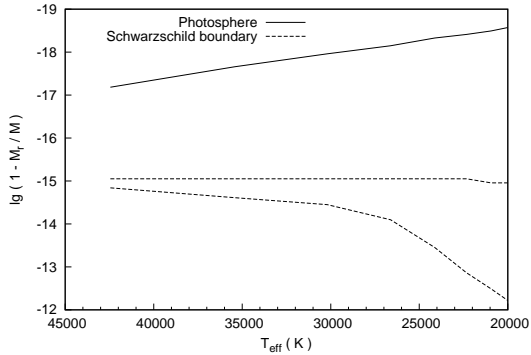}
}
\caption{\label{crml2}Similar to Figure \ref{crm}, but $l=2H_{\rm P}$.}
\end{figure}

Furthermore, a more efficient convection can change the 
deepening of the inner boundary of the He convection during the 
white dwarf's evolution, and thus change the transition temperature. In the MLT 
(mixing-length theory), the mixing-length $l$ represents the efficiency 
of convective heat transfer. We considered a series of models, in which the 
mixing-length is set to $2H_{\rm P}$, which is $2$ times larger than before. 
Our calculations show that the deepening of the convection zone do occur at a 
higher effective temperature. As shown in Fig. \ref{crml2}, the convection 
zone of a model with $M_{\rm H}=10^{-15}~M_{\odot}$ deepening at 
$T_{\rm eff}\sim30,000~{\rm K}$ (compared with Figure \ref{crm}). For this reason 
Eq. (\ref{eqtr}) will be satisfied earlier and it will lead to a change of $T_{\rm tr}$ 
(see Table \ref{temptr}). However, the variation of $T_{\rm tr}$ is relatively small, 
especially around $30,000~{\rm K}$. We therefore believe that the efficiency of the 
convection will not significantly affect the results.

\begin{figure}[ht]
\begin{center}
\includegraphics{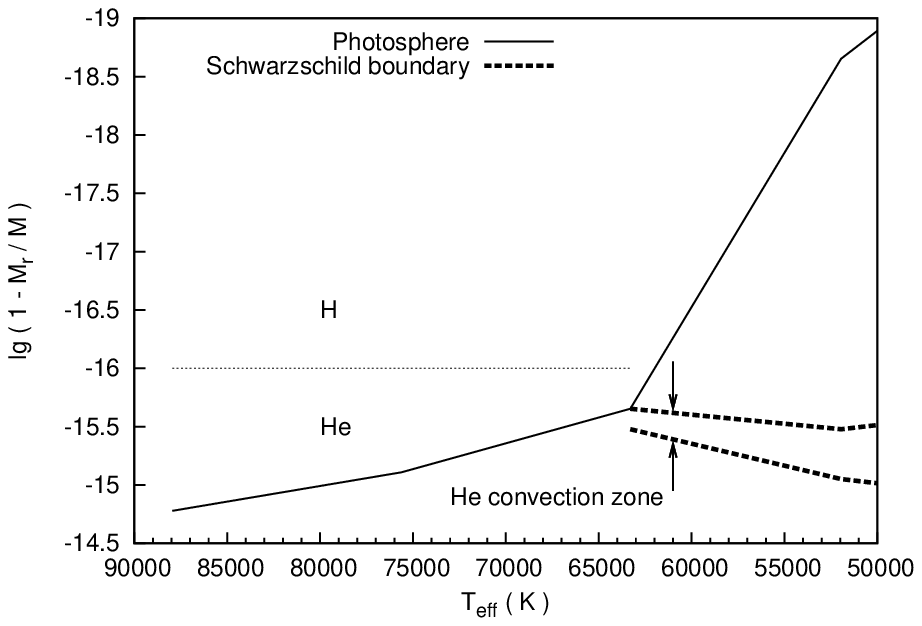}
\caption{A schematic representation of the formation of a DO white dwarf from a DA star
with $M_{\rm H}\sim10^{-16}~M_{\odot}$.}
\label{fmtdo}
\end{center}
\end{figure}

The above discussions imply that DB white dwarfs are likely born
from DA white dwarfs because of the convective mixing. Is it
possible that a DO white dwarf can evolve into a DA star? Our
calculations show that if a DA white dwarf with a sufficiently thin
hydrogen layer of $M_{\rm H}\sim10^{-16}~M_{\odot}$, this
transformation is possible. As shown in Figure \ref{fmtdo}, during
the early time of the evolution, the model's
 $T_{\rm eff}$ is very high and hydrogen is completely ionized in its atmosphere.
The photosphere lies deeply in the helium layer and thus helium is visible, resulting in
the white dwarf appearing as a DO star. When the white dwarf model cools down
to $T_{\rm eff}\sim63,000~{\rm K}$, the photosphere rises to the hydrogen layer, but at
the same time convection appears in the He II/III ionization zone. Because of the thin
hydrogen layer, the convective overshooting can possibly reach to the photosphere and
Eq. (\ref{eqtr}) can easily be satisfied. The convective motion in the helium layer will
dilute the hydrogen layer, so the white dwarf is still a DO star.

\begin{figure}[ht]
\begin{center}
\includegraphics{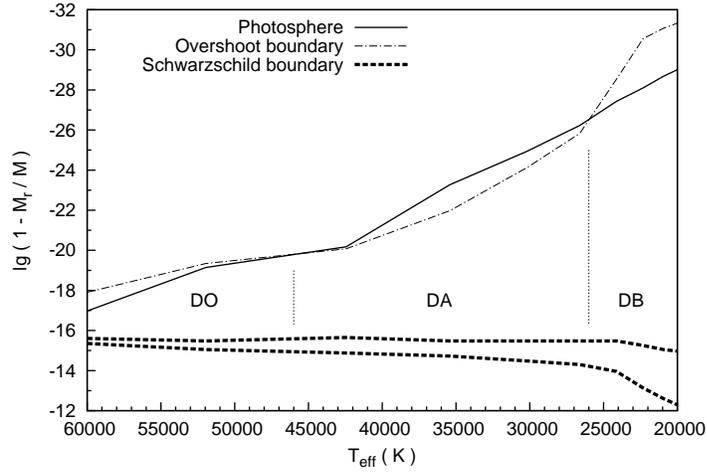}
\caption{A schematic representation of the spectral evolution of a whte dwarf model with
$M_{\rm H}\sim10^{-16}~M_{\odot}$. The dash-dotted line corresponds to the boundary
of the overshooting region. }
\label{h16}
\end{center}
\end{figure}

When the white dwarf cools down below $60,000~{\rm K}$, as shown in
Figure \ref{h16}, the location of the photosphere will quickly move
up toward the stellar surface. The extension of the convective
motion can not go so far to reach the photosphere. Therefore the
convective mixing of helium is invisible and hydrogen will
re-accumulate in the atmosphere, making the DO white dwarf
transforming into a DA star. This result is similar to the
Shibahashi's assumption (Shibahashi \cite{Sa}, \cite{Sb}). According
to the observational data, there are no DO white dwarfs below
$T_{\rm eff}\sim45,000~{\rm K}$. This fact allows us to adjust the
overshooting length to let our model change its spectral type at
$T_{\rm eff}\sim45,000~{\rm K}$. We can assume reasonably that the
overshooting length is proportional to the dimension of the
convection zone. In practice, we choose the overshooting length to
be the length of convection zone (being approximately equal to
$H_{\rm P}$) plus $0.375H_{\rm P}$. Our numerical result shows that
at an effective temperature ($T_{\rm eff}\sim26,000~{\rm K}$) the
convection zone has developed thick enough that the convective
overshooting can reach again above the photosphere. Then the
hydrogen layer will be mixed with the helium layer and the white
dwarf will become a DB star.

There is another possibility for DO white dwarfs transforming 
into DA stars in the literature. It is probably that the progenitors of 
DO white dwarfs have undergone a born-again phase and burnt most of the 
hydrogen envelope (Althaus et al. \cite{A}). The mass loss due to stellar 
wind during the hot white dwarf stage is likely to throw away the 
superficial hydrogen and prevents the gravitational settling (Unglaub \& Bues \cite{UB}). 
When the white dwarfs cool down to $T_{\rm eff}\sim45,000~{\rm K}$, hydrogen previously
left in the internal layer is able to float to the stellar surface and DA white dwarfs 
are formed.

\section{Conclusions}
\label{sect:conclusion}

\begin{figure}[ht]
\begin{center}
\includegraphics{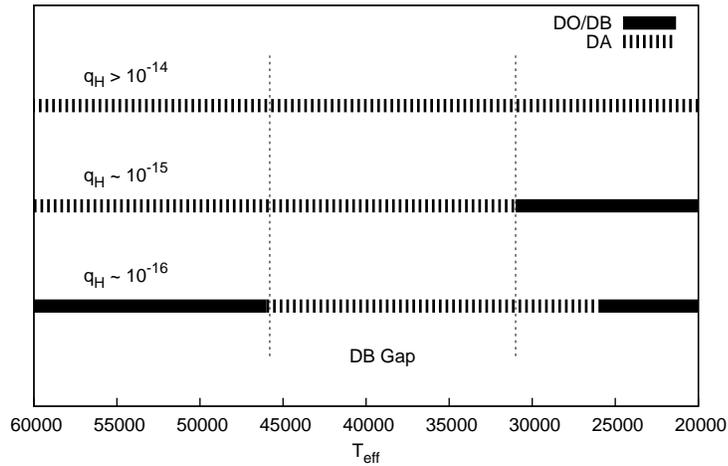}
\caption{A schematic representation of the DB gap. $q_{H}=M_{H}/M_{\odot}$.}
\label{abo}
\end{center}
\end{figure}

From the above investigations we have found that the DB gap could be explained as
a consequence of the convective mixing in white dwarfs. DA white dwarfs
with $M_{\rm H}/M_{\odot}\sim10^{-16}$ have opportunities to transform into
DO white dwarfs at $T_{\rm eff}\gtrsim46,000~{\rm K}$ or DB white dwarfs
at $T_{\rm eff}\lesssim26,000~{\rm K}$, respectively. DA white dwarfs
with $M_{\rm H}/M_{\odot}\sim10^{-15}$ will can transform into DB stars
below $T_{\rm eff}\sim31,000~{\rm K}$. White dwarfs with $M_{\rm H}$ greater
than $10^{-14}~M_{\odot}$ always appear as DA stars
at $T_{\rm eff}\gtrsim18,000~{\rm K}$. It is obvious that in the effective temperature
range between $46,000\lesssim T_{\rm eff}\lesssim31,000~{\rm K}$ almost all of white
dwarfs have the DA spectral type, as shown in Figure \ref{abo}. This scenario
substantially coincides with the observation. We can also estimate that the hydrogen
mass is $M_{\rm H}/M_{\odot}\sim10^{-16}$ for the DO white dwarfs and
is $M_{\rm H}/M_{\odot}\sim10^{-15}$ for the hot DB white
dwarfs ($T_{\rm eff}>20,000~{\rm K}$).

Based on our numerical results, the convective overshooting plays a crucial role in
the formation of the so-called DB gap, through the convective mixing effect. It allows
the convective motion penetrating into the hydrogen layer and makes helium in the deep
stellar interior being observable on the stellar photosphere. The overshooting length is
an important parameter of the model. According to our results, the overshooting length
should be proportional to the thickness of the convection zone, which gives better
agreement between the model results and observations.

The hydrogen mass $M_{\rm H}$ is another important parameter, which is used as a
criterion for deciding when helium is dominant in the atmosphere of the white dwarf. It
determine decisively the critical effective temperature for the white dwarf changing its
spectral type.

\begin{acknowledgements}
We thank Q. S. Zhang for many valuable discussions.
This work is supported by the National Key Fundamental Research Project through grant 2007CB815406.
\end{acknowledgements}

\label{lastpage}

\end{document}